# Impact of capping agent on structural and optical properties of ZnS nanoparticles

[1]Samiran Mandal*, [2]Sk Irsad Ali, [3]Subhamay Pramanik and [2]Atis Chandra Mandal

[1]Government General Degree College At Pedong,Pedong,Kalimpong-734311, West Bengal, India

[2]Department of Physics, University of Burdwan, Golapbag, Burdwan 713104, West Bengal, India

[3]School of Nanoscience and Technology, IIT Kharagpur-721302, West Bengal, India

*Email : samiranmandal1@gmail.com

**Abstract:**

Nanocrystalline samples of pristine capped and uncapped zinc sulphide were synthesized via the sol-gel technique. The nanocrystallinity of the samples were confirmed by the X-ray diffraction technique, where size of the particle size decreases with the increasing of mol. concentration (x = 0.00, 0.02, 0.03, 0.04 Mol). of capping agent sodium dodecyle sulphate. The obtained crystallite sizes were found to be in the range 4.6 nm to 2.7 nm respectively. The optical band gaps of the samples were estimated by using ultra-violet visible spectroscopic techniques and the band gap values were in the range 3.8 eV to 4.4 eV. All the samples showed quantum confinement behavior compared to bulk sample. Fluorescence (FL) spectra showed three emission peaks at the emission wavelengths around 434 nm, 520 nm, 545 nm, 628 nm, and 694 nm. The FL intensities were proportional to the concentration of capping agent.

## 1. Introduction

Based on the significant dependence on particle size, the synthesis of nanoparticles (NPs) has been a flourishing field in recent years. Because of the wide range in bandgap, II-VI class inorganic semiconductor materials such as ZnS, CdS, CdSe, and ZnSe have shown to be versatile materials for use in optoelectronic devices. ZnS has received a lot of interest because of its uses in efficient phosphors, flat panel displays, cathode ray tubes, solar cells, and so on. Semiconductor NPs are a type of material that combines molecular and bulk characteristics. They have received a lot of interest in recent years due to their new features which result from the quantum confinement effect [1-4]. When the diameters of the nanoparticles are comparable to the Bohr excitonic radius of those materials, the quantum confinement effect affects the electronic structure of the nanocrystals. When the particle radius approaches the excitonic Bohr radius, the band gap energy expands, resulting in a blue shift in the band gap, emission spectra, and so on. Surface states, on the other hand, will play a more significant part in NPs owing to their large surface-to-volume ratio with decreasing particle size (surface effects). In the case of semiconductor NPs, recombination can be either radiative or nonradiative. These size-dependent optical features have a wide range of applications; the ability to tune the properties of NPs by regulating their size may be advantageous in developing novel composite materials with optimised qualities for a variety of applications including solar energy conversion, light-emitting devices, chemical or biological sensors, and photo catalysis [5-9].

Thus the size of the NPs is the key parameter of its efficacy. With the reduction in size of NPs, the samples get agglomerated due to high reactive surface atoms of the same and lost its fundamental characteristics; especially the quantum dots are very much susceptible to this nature. To prevent this difficulty it is very much necessary to encapsulate the atoms or molecules by using surfactant or capping agent such that no further agglomeration can take place.

In the present investigation we have used bottom up or chemical approach. ZnS NPs are synthesised using simple sol-gel technique where sodium dodecyle sulphate was used as capping agent. The material synthesis procedure of ZnS NPs has been described with uncapped and different varying mol concentration of sodium dodecyle sulphate. The samples were characterized by different spectroscopic method like X-ray diffraction(XRD), Optical absorption (OA) by Ultra violet visible spectroscopy (UV-Vis) and Fluorescence (FL) to find the structural correlation of the same.

## 2. Experimental Procedure:

### 2.1 Synthesis of ZnS NPs.

ZnS NPs were synthesized by chemical method especially by sol-gel method. Here we used analytic grade chemical reagents like 99% zinc nitrate hexahydrate [$Zn(NO_3)_2, 6H_2O$], sodium sulphide nonahydrate [$Na_2S, 9H_2O$], Sodium hydroxide [$NaOH$] and as capping agent sodium dodecyle sulphate [$CH_3(CH_2)_{11}OSO_3Na$]. At room temperature individual 1 mol solutions of zinc nitrate was mixed with different concentration (x = 0.00 M, 0.02M, 0.03M and 0.04M) solution of capping agent. The mixtures were stirred for 30 minutes by magnetic stirrer during stirring 1 mol individual solution of sodium sulphide were drop wise added to the reaction mixture. Milky white solutions were formed. Further, 1 (N) NaOH solution drop by drop mixed with the white complex solution until it becomes alkaline. The stirring continues at 400 rpm for 3 hours more. After 3 hours stirring the colloidal solution was left for 12 hour to precipitate. The white gel like precipitates were filtered and washed several times to get rid of soluble impurities. Later the individual white precipitate as dried in a oven at $100^{O}C$ for 5 hours each. After fine grinding these samples were individually annealed at temperature $250^{O}C$ for 3 hours and were taken for different characterization.

The chemical reaction equation to the synthesized ZnS NPs defined as:

$Zn(NO_3)_2, 6H_2O + Na_2S, 9H_2O \longrightarrow ZnS + 2NaNO_3 + 15H_2O$

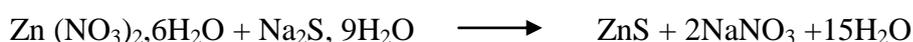

## 2.2. Characterization techniques

The XRD spectra recorded with the help of a Bruker D8 Advance diffractometer having incident radiation CuKα (λ = 0.15418 nm), the operating parameters were at 40mA and 40 kV. The angular range (2θ) of diffractometer is $10-80^0$ having scanning rate $2^0$/minute. Shimadzu, UV-1800 double-beam spectrophotometer was utilised for optical absorption (OA) analysis, which utilised sonicated homogenised de-ionized water solutions of very small amounts of ZnS nanoparticulate samples. The Perkin Elmer LS 55 fluorescence (FL) spectrometer with an excitation wave length of 350 nm was used to study the FL emission spectra.

## 3. Result and Discussion

### 3.1 XRD Analysis:

Figure 1 depicts the XRD patterns of synthesised ZnS NPs. The diffraction pattern for each sample is connected with three planes with miller indices of (1 0 16), (1 1 0) and (0 2 16) and the ZnS NPs have a rhombohedral shape. (JCPDS No. 89-2426).

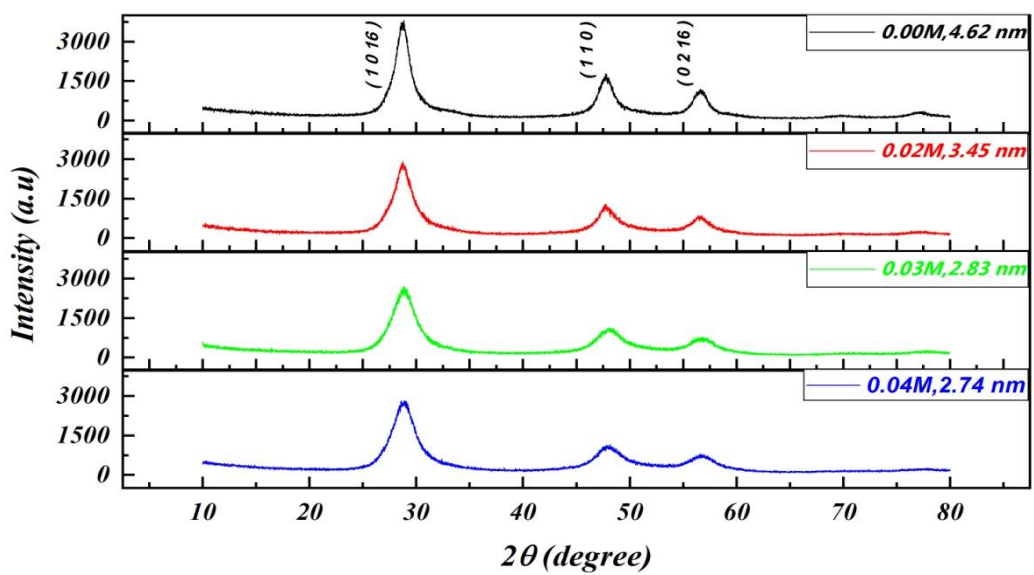

**Fig. 1:** (colour online) XRD patterns of ZnS NPs at different capping agent concentration.

Using the Debeye-Scherrer formula [9-12], the average particle size was calculated by taking the full-width at half-maximum (FWHM) of all three peaks.

$$D = \frac{0.94\lambda}{\beta \cos\theta} \qquad (1)$$

Where D is the particle's mean diameter, is the wavelength of X-ray radiation (1.5418A), β is the FWHM (rad.) of the XRD peaks, and θ is the angle of diffraction. The FWHM was calculated using the software Origin (Version 8.5) and Gaussian fitting. Table 1 shows the particle sizes found, which are in good accord with previous findings (13, 14, 15).

**Table.1:** Variation of nanocrystallite sizes with varying concentration of capping agent

| Concentration of Capping Agent (Mol.) | Crystallite sizes(nm) |
|---|---|
| 0.00 | 4.62 |
| 0.02 | 3.45 |
| 0.03 | 2.83 |
| 0.04 | 2.74 |

From the table 1 it is evident as the molar concentration of capping agent increased the sizes of the obtained NPs were reduced (13, 14, 15). This indicated the increase of capping agent concentration successfully prevent the agglomeration of the synthesized particles by encapsulating them to smaller individuals.

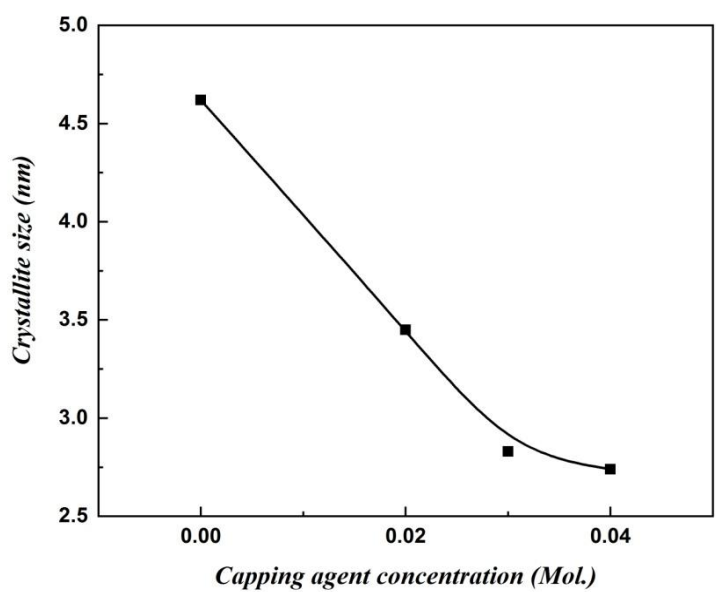

**Fig.2:** Variation of crystallite size with molar concentration of capping agent

**3.2 OA absorption measurements:**

To investigate the behaviour of semiconductor nanocrystals requires an understanding of optical absorption. The band gap--the energy gap between the filled valence band and the empty conduction band is essential characteristic of semiconductors. Electron optical stimulation across the band gap is strongly permitted, resulting in a rapid increase in absorption at the wavelength corresponding to the band gap energy. The optical absorption edge is an element of the optical spectrum. The optical absorption spectra of ZnS NPs in the 800 nm-200 nm region are shown in Fig. (3). the spectra are featureless as no absorption occurs in the visible band (800 nm-390 nm). Absorption edges were achieved in the UV region at shorter wavelengths of 286 nm, 306 nm, 314 nm, and 326 nm for the capping agent concentration 0.00M, 0.02M, 0.003M and 0.04M respectively.

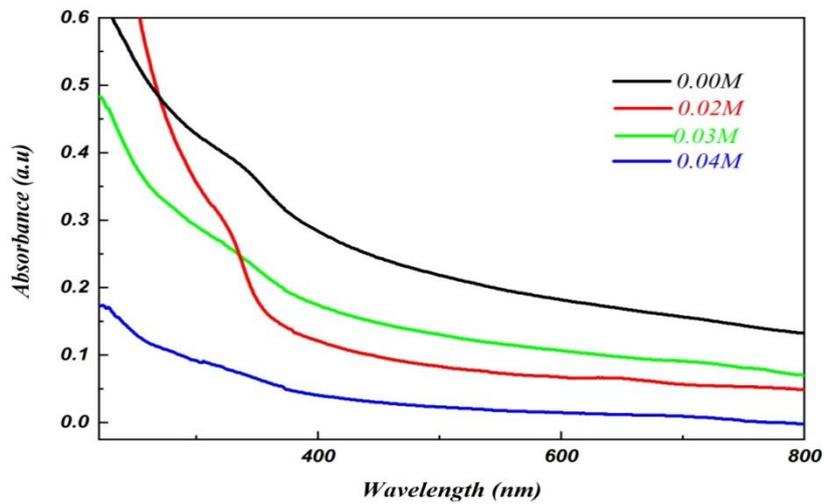

**Fig.3:** (colour online).UV-Vis absorption spectra of ZnS NPs of different capping agent concentration

This clearly illustrates that when the capping agent concentration increases, the absorption edge shifts towards shorter wavelengths. The observed blue shift in the absorption edge is a reflection of the increased band gap caused by the quantum confinement effect.

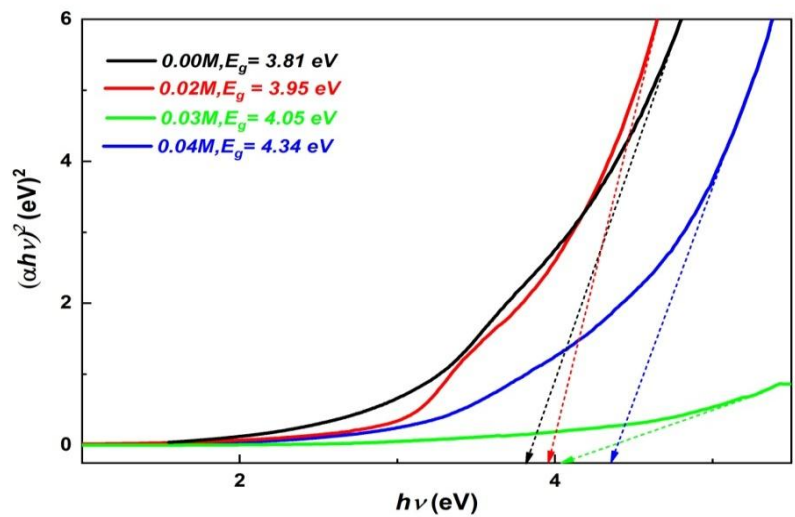

**Fig.4:** (colour online) Tauc plot of UV-Vis absorption spectra of ZnS NPs for different capping agent concentration

As expected, there are no absorption peaks in the visible region. The optical band gap of prepared ZnS NPs may be calculated using the basic absorption principle, which involves electrons being stimulated from the valence band to the conduction band. Applying the Tauc equation (16, 17) the optical band gap of the NPs is estimated from the OA absorption study.

$$\alpha v h = A(h\nu - E_g)^n \qquad (2)$$

In this equation, $h\nu$ is energy of incident photon, $\alpha$ is the absorption coefficient, A is a constant, and $E_g$ is the optical band gap energy of the material. The value of the exponent (n) varies depending on the nature of the transition, for example, the value of n=1/2 for direct transitions and 2 for permissible indirect transitions. The direct optical band gap of ZnS NPs is estimated from the Tauc plot by plotting $(\alpha h\nu)^2$ Vs. $h\nu$ and then extrapolating the tangent of the linear portion of the curve to the $(\alpha h\nu)^2 = 0$ axis, as illustrated in Fig. 4.

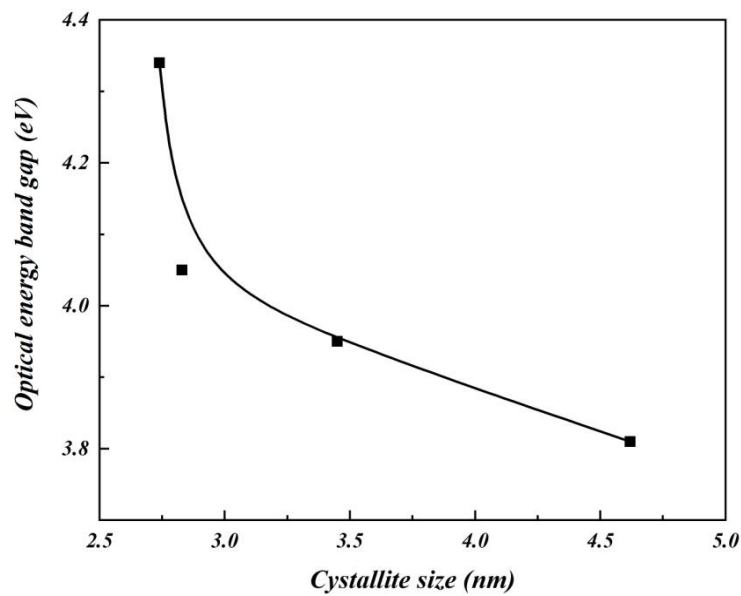

**Fig.5:** (colour online) variation of optical energy band gap ZnS NPs for different capping agent concentration with respect to crystallite size.

According to the comparison investigation, as particle size drops from 4.62 nm to 2.74 nm, the acquired value of optical band edge shifts blue from 3.81 eV to 4.34 eV (fig.5). It demonstrates the onset of quantum size effects (18, 19) in which electron-hole confinement within a small crystallite size causes an increase in band gap energy, showing a shift in the absorption spectra of ZnS NPs to the lower wavelength side. These findings are consistent with previous publications (20, 21) Figure 5 depicts the energy band gap vs. particle size. This normal change in band gap energy caused by NP size variation might be explained by the electron confinement process or by the introduction of defects states within the band gap.

The size of the NPs also can be determined using the following Brus equation (22) for quantum dots

$$E_g^{nano} = E_g^{bulk} + \frac{h^2 \pi^2}{2r^2}\left(\frac{1}{m_h^*} + \frac{1}{m_e^*}\right) - \frac{1.8e^2}{\epsilon r} \qquad (3)$$

Where $E_g^{nano}$ and $E_g^{bulk}$ are the optical band gap of NPs and corresponding bulk materials. $m_h^*$ and $m_e^*$ are the effective mass of hole and electron are equal to $0.61 m_e$ and $0.41 m_e$ respectively. $m_e$ is the free mass of electron. r is the diameter of the NPs, e is the electronic charge and $\epsilon$ is the dielectric constant of the material. The sizes of the NPs obtained by the above formula are nearly same order of magnitude obtained from the XRD data.

### 3.3 Fluorescence spectra analysis:

Fluorescence studies provide information about defect states that contribute to the sample's radiative de-excitation. FL investigations demonstrate that the defect states in nanocrystals can alter or increase in density. The FL spectra of uncapped and capped ZnS NPs samples with varied capping agent concentration are shown in Fig. (6). The figure shows that FL peaks are found for all the samples at the wavelengths 434 nm, 520 nm, 545 nm, 628 nm, and 694 nm. The limitation of FL spectrophotometer did not permit to record the full intensity at the wavelength 694 nm. This is also justifies ZnS as a highly luminescent material. The FL emission wavelength is much longer than the excitation wavelength. (350 nm) This suggests that impurities or surface states are involved in the FL phenomena.

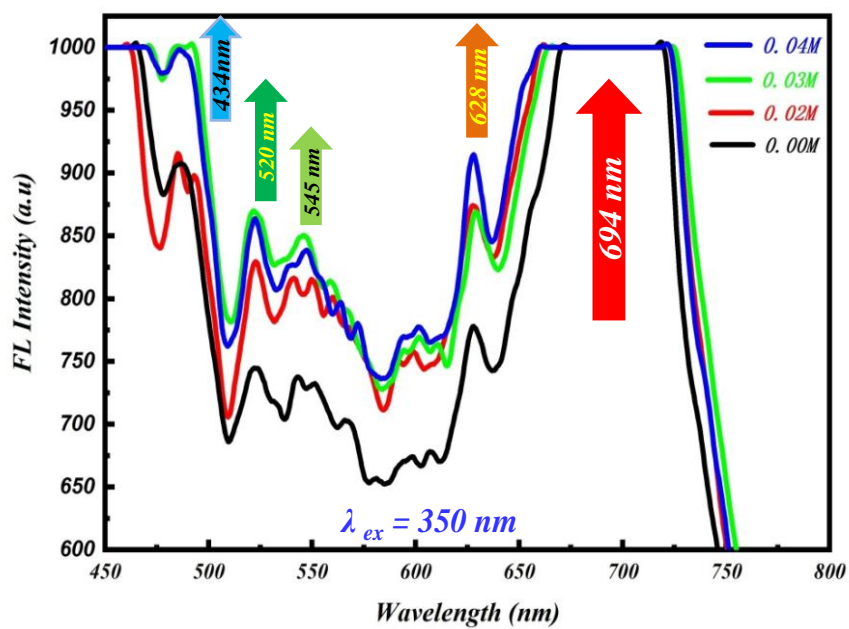

**Fig.6:** (Colour online) Variation of fluorescence intensities of different capped and uncapped ZnS NPs.

The FL spectra of ZnS nanoparticles with varying capping agent concentrations show that as particle size is decreased, the emission gets more intense and slightly shifted to a shorter wavelength. According to photoluminescence spectra, uncapped ZnS NPs have a broad emission peak at 434 nm that originates from the host ZnS. This blue emission is caused by a self-activated center, which is thought to be produced between a Zn vacancy and a shallow donor connected with a Sulphur vacancy (22). The green emission peak near 550 nm is attributed to a deep-level transition involving self-activated point defects such Zn and S vacancies (19, 23). The other emission peaks around the 628 nm and 694 nm might be due to deep level of transition related to $Zn^{+2}$ and $S^{-2}$ vacancy type defects. The emission intensities of capped ZnS NPs are higher than that of uncapped ZnS. This is to be expected since in the absence of a capping agent, uncontrolled particle nucleation and growth took place, leading to the creation of defect states. However, improved luminescence was seen in capped NPs due to surface modification by capping agents, which had the impact of minimizing surface defects, as well as energy transfer from chemisorbed capped molecules to interstitial sites and vacancy centers. The enhancement of luminescence intensity with decreasing particle size has been attributed to the quantum size effect (24, 25). The valence band edge shifts downward as particle size decreases. As a result, the released photon has a larger energy, resulting in a photoluminescence peak at a shorter wavelength (19).

## 4. Conclusion:

The capped and uncapped ZnS NPs were successfully synthesized by sol-gel method and are characterized by different spectroscopic technique such as XRD, UV-Vis, and Fluorescence. With the increase of capping agent concentration the size of ZnS NPs are found to be decreased indicating successful prevention from the particle agglomeration. A broad tunability in the optical band gap energy from 4.34eV to 3.81eV is found from the varied sizes of ZnS NPs which was obtained by adjusting initial concentration of capping agent. The optical energy band gap of the said NPs samples have shown onset of strong quantum confinement effect compared to its bulk counterpart. These particles exhibit significant fluorescence; however, the existence of defects has a strong influence on it. Because of these defects ZnS NPs are more reactive than the bulk particles, they may be beneficial for surface reactions that affect optical characteristics as the size of NPs decreased. This give rise to more and more intense peak at different emission wavelengths compared to uncapped one and indicates the formation of surface defects owing to $Zn^{+2}$ and $S^{-2}$ vacancy within the samples.

**Acknowledgement:**

The author grateful to Prof. Atis Chandra Mandal Department of Physics, The University of Burdwan for providing necessary facility to carry out the experiments.